\documentclass{pasj00}

\begin{document}
\SetRunningHead{K. Motohara et al.}{CISCO: Cooled Infrared Spectrograph
and Camera for OHS}
\Received{2001/11/19}
\Accepted{2002/02/20}

\title{CISCO: Cooled Infrared Spectrograph and Camera for OHS on the
Subaru Telescope}

\author{
  Kentaro \textsc{Motohara}\altaffilmark{1,2,3},
  Fumihide \textsc{Iwamuro}\altaffilmark{3},
  Toshinori \textsc{Maihara}\altaffilmark{3,4}\\
  Shin \textsc{Oya}\altaffilmark{3,5},
  Hiroyuki \textsc{Tsukamoto}\altaffilmark{3},
  Masatoshi \textsc{Imanishi}\altaffilmark{3,6},\\
  Hiroshi \textsc{Terada}\altaffilmark{2,3},
  Miwa \textsc{Goto}\altaffilmark{2,3},
  Jun'ichi \textsc{Iwai}\altaffilmark{3},
  Hirohisa \textsc{Tanabe}\altaffilmark{3},\\
  Ryuji \textsc{Hata}\altaffilmark{3},
  Tomoyuki \textsc{Taguchi}\altaffilmark{3}, and
  Takashi \textsc{Harashima}\altaffilmark{3},
}

\altaffiltext{1}{Institute of Astronomy, The University of Tokyo, Mitaka, Tokyo 181-0015}
\email{kmotohara@ioa.s.u-tokyo.ac.jp}
\altaffiltext{2}{Subaru Telescope, National Astronomical Observatory of Japan, 650 North A'ohoku Place, Hilo, HI 96720, USA}
\altaffiltext{3}{Department of Physics, Kyoto University, Kitashirakawa, Kyoto
606-8502}
\altaffiltext{4}{Department of Astronomy, Kyoto University,
Kitashirakawa, Kyoto 606-8502}
\altaffiltext{5}{Communications Research Laboratory, Koganei, Tokyo 184-8975}
\altaffiltext{6}{Optical and Infrared Astronomy Division, National
Astronomical Observatory of Japan, Mitaka, Tokyo 181-8588}

\maketitle

\begin{abstract}
This paper describes a Cooled Infrared Spectrograph and Camera for OHS 
(CISCO), mounted on the Nasmyth focus of the Subaru telescope.
It is primarily designed as a back-end camera of the OH-Airglow Suppressor
(OHS), and is also used as
 an independent, general-purpose  near-infrared camera/spectrograph.
CISCO is based on a single 1024$\times$1024 format HgCdTe HAWAII array
detector, and is capable of either wide-field imaging of
 \timeform{1.8'}$\times$\timeform{1.8'} field-of-view or low-resolution
 spectroscopy from 0.9 to 2.4 $\mu$m. 
The limiting magnitudes measured during test
 observations were found to be $J=23.5$ mag and $K^{\prime}=22.4$ mag 
(imaging, \timeform{1''} aperture, $\rm S/N=5$, 1 hr exposure).
\end{abstract}

\section{Introduction}
Recent progress in large-format array detectors for infrared
wavelengths has enabled us to obtain spectacular images of forming stars,
nuclei of active galaxies and the rest-optical distant universe
with performance comparable to that of optical CCDs. 
Together with the current 8--10m telescopes, we now have
powerful tools to explore the universe.

The Cooled Infrared Spectrograph and Camera for OHS (CISCO) is one of such
 instruments, exploiting a single 1024$\times$1024 format array 
 detector (HAWAII) and installed on the Nasmyth focus of the Subaru
telescope. 
It was originally designed as a back-end camera
for an OH-airglow suppressor (OHS: Iwamuro et al. 2001), 
and was also intended to be used as a general-use infrared camera
capable of wide-field (\timeform{1.8'}$\times$\timeform{1.8'}
FOV) imaging as well as long-slit low-resolution spectroscopy from 
0.9 to 2.4 \micron. 

We describe the overall design of the instrument in section 2, 
report on the performance of the HAWAII array detector
in section 3 and give the total system performance in section 4.

\section{Design of the Instrument}

\subsection{Overview}
\begin{figure}
  \begin{center}
    \FigureFile(150mm,80mm){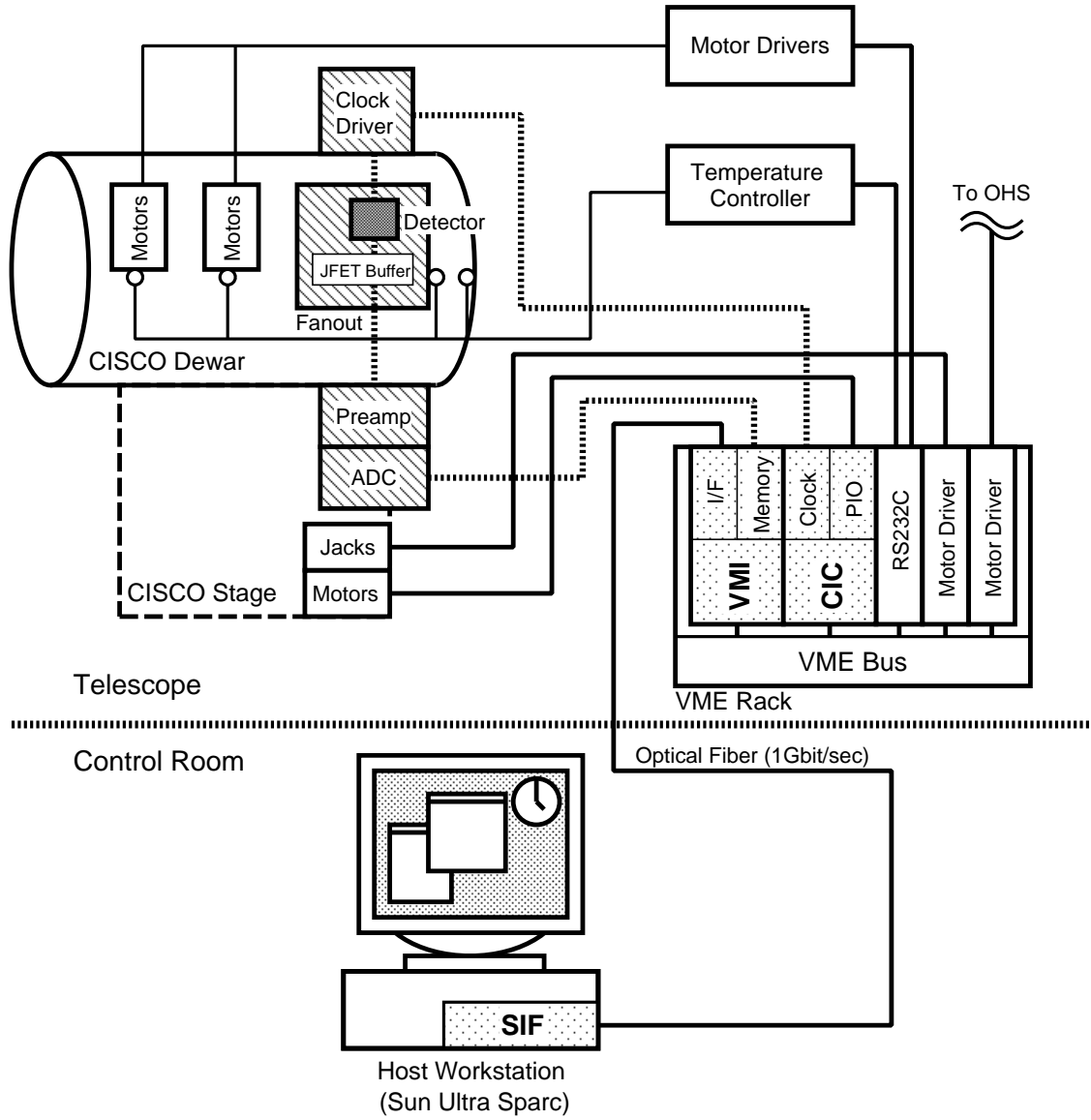}
  \end{center}
  \caption{Block diagram of the control system of CISCO.}\label{fig:layout}
\end{figure}

Let us first describe the overall configuration of CISCO.
Figure \ref{fig:layout} shows a block diagram of the whole system of
CISCO. It is controlled by a workstation placed at the 
control building. This workstation is connected to a VME board computer
system (Messia III: Sekiguchi et al. 1992) by optical fibers, which
handles the data-acquisition system, all motors including those of OHS, and a
temperature controller. 

\begin{figure}
  \begin{center}
    \FigureFile(170mm,80mm){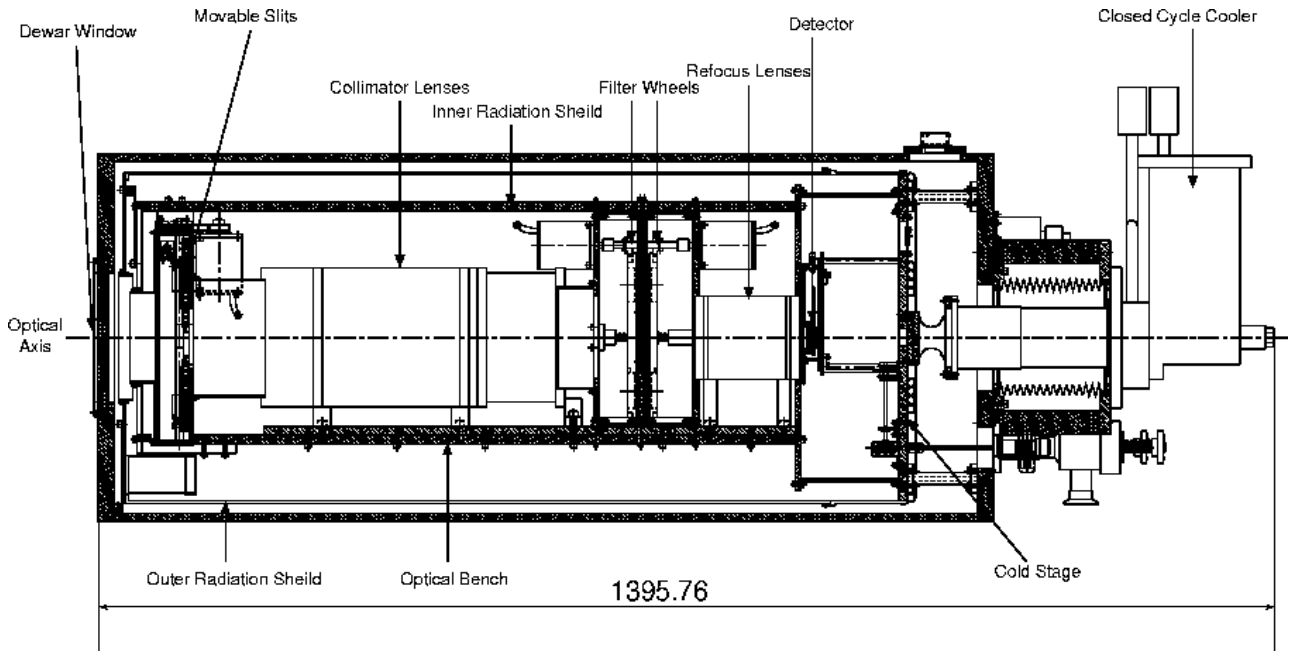}
  \end{center}
  \caption{Schematic view of the cross-section of the cryostat of CISCO.}\label{fig:cryostat}
\end{figure}

Figure \ref{fig:cryostat} shows a cross-sectional view of the 
cryostat of CISCO. 
 The entrance slits of the optics consists of two pairs of 
slit blades, enabling various slit configurations from a pin-hole 
to a wide-field square aperture, including a long slit mode. 
 Two filter wheels are placed in the collimated beam and 
contain nine filters, three grisms and one prism. The imaging mode and
the spectroscopy mode are switched by changing the filters/grisms and
opening/closing the slit. 
The whole interior is cooled by a closed-cycle cooler.

The cryostat of CISCO is mounted on an adjustable stage and installed 
either at the end of the optics of OHS or at the
Nasmyth focus of the telescope.

\subsection{Cryogenics}
The enclosure of the cryostat is a cylinder made of aluminum-alloy, 1000
mm in length and 400 mm in diameter, which was fabricated and assembled
by Infrared Laboratories, Inc. 
A cold stage is located on the right end of the cryostat in figure
\ref{fig:cryostat}, fixed to the cryostat by four poles made of glass
epoxy resin. 
An optical bench and a detector-cassette are mounted to this 
cold stage.

The whole structure on the cold stage is covered by an outer radiation
shield to avoid heat flow from outside, and the optical
bench is also covered by an inner radiation shield to obstruct
residual heat radiation.
The outer shield is connected to the left edge of the cryostat by
supports made of glass epoxy resin;
all of the interior is held rigidly.

We use a model 1050 cold head and a model 1020 air-cooled compressor
manufactured by CTI Cryogenics Inc. for cooling.
The cold head is fixed to the cryostat by 
a bellows and four poles with rubber washers to absorb vibration of
the cold head. 
The cold head and the cold stage are connected by a bundle of thin copper
sheets to insulate the vibration.
Additional copper tube for flowing 
liquid nitrogen to accelerate the cooling speed is located
under the cold stage.

Thermometers made of silicon diode are located at the slit box, the filter
wheels, the detector cassette and the cold head.
We use a temperature monitor/controller manufactured by Lakeshore Inc. for
temperature data acquisition, which also controls the temperature of the
detector cassette using a heater.
Under normal operation, the temperature at the detector cassette is 
set to 77 K.

It takes about 70 hr to cool down the whole system using only a
mechanical cooler.
The normal operation temperature is 54 K at the cold head, 57 K at the
filter wheels, and 59 K at the slit box.
The vacuum pressure reaches as low as $1\times 10^{-7}$ torr.

\subsection{Optics}
\begin{figure}
  \begin{center}
    \FigureFile(130mm,80mm){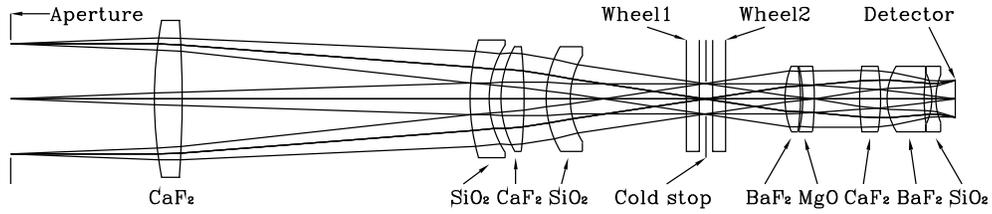}
  \end{center}
  \caption{Optical layout of CISCO.}\label{fig:optics}
\end{figure}

The optical design is shown in figure \ref{fig:optics},
and the parameters of the optical elements are listed in table \ref{tab:opt}.
The surfaces of all the lenses are spherical.
These lenses are installed in lens holders made of invar and fixed on 
the optical bench.

We have two choices for the secondary mirror of the telescope:
one is an infrared secondary mirror; the other is a Nasmyth-optical
secondary mirror.  
The focal ratio of a ray from the telescope is f/13.6 for the infrared
secondary and f/12.7 for the Nasmyth-optical.

At the focal plane of the telescope, two movable slit-units
consisting of two blades are set 
perpendicular to each other. 
We obtain a wide-field image when the both units are
opened, a long slit image when one is closed, and a pinhole image smaller than
a single pixel if both units are closed.

The ray is next collimated and lead to 
a Ly\^ot stop.
Two filter wheels are arranged on both sides of the stop.
Each wheel has eight apertures, one being left to be a through hole.  
5 broad-band filters, 4 narrow-band filters, 3 grisms and 1 prism 
shown in table \ref{tab:filter} are installed here.
As for the broad-band filters, we employed that of Mauna Kea Observatories
NIR Filter Set (Simons, Tokunaga 2002; Tokunaga et al. 2002).

\begin{figure}
  \begin{center}
   \FigureFile(150mm,80mm){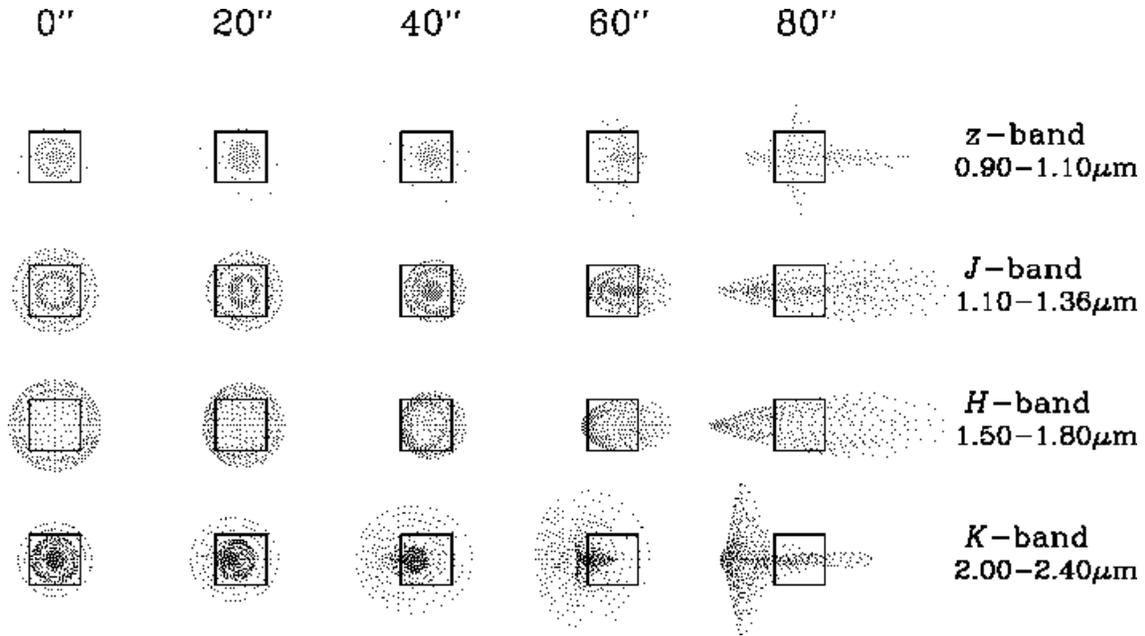}
  \end{center}
  \caption{Spot diagrams of CISCO. Solid squares represent the size of the pixel.}\label{fig:spotdiagram}
\end{figure}

The collimated ray is then re-focused on the array detector.
Figure \ref{fig:spotdiagram} shows spot diagrams of a
refocused image.
Throughout the wavelength coverage of CISCO, the image distortion is less
than 2 pixels and the image size is smaller than 2 pixels for a full frame.

The pixel scale varies according to the choice of the secondary mirror
of the telescope. It is \timeform{0.105''}/pix with the infrared
secondary and \timeform{0.111''}/pix with Nasmyth-optical secondary, and 
the field of view is \timeform{108''}$\times$\timeform{108''} and
\timeform{114''}$\times$\timeform{114''}, respectively.

The wavelength resolution of spectroscopy mode is
$\lambda/\Delta\lambda=230$ for the $zJ$ grism, 200 for the $JH$
and 250 for the $K$ using the slit width of 9.5 pixels
(this corresponds to \timeform{1.0''} when using the infrared secondary).

\subsection{Mechanics}
We use four stepper motors manufactured by Phytron Co. to drive 
the two filter wheels and the two slit-units.
They are installed inside the cryostat, and cooled down below 
60K.

The motors are connected to drivers outside the cryostat, 
controlled from the host workstation via an RS-232C board in the 
VME rack. 

\subsection{CISCO Stage}
\begin{figure}
  \begin{center}
   \FigureFile(150mm,80mm){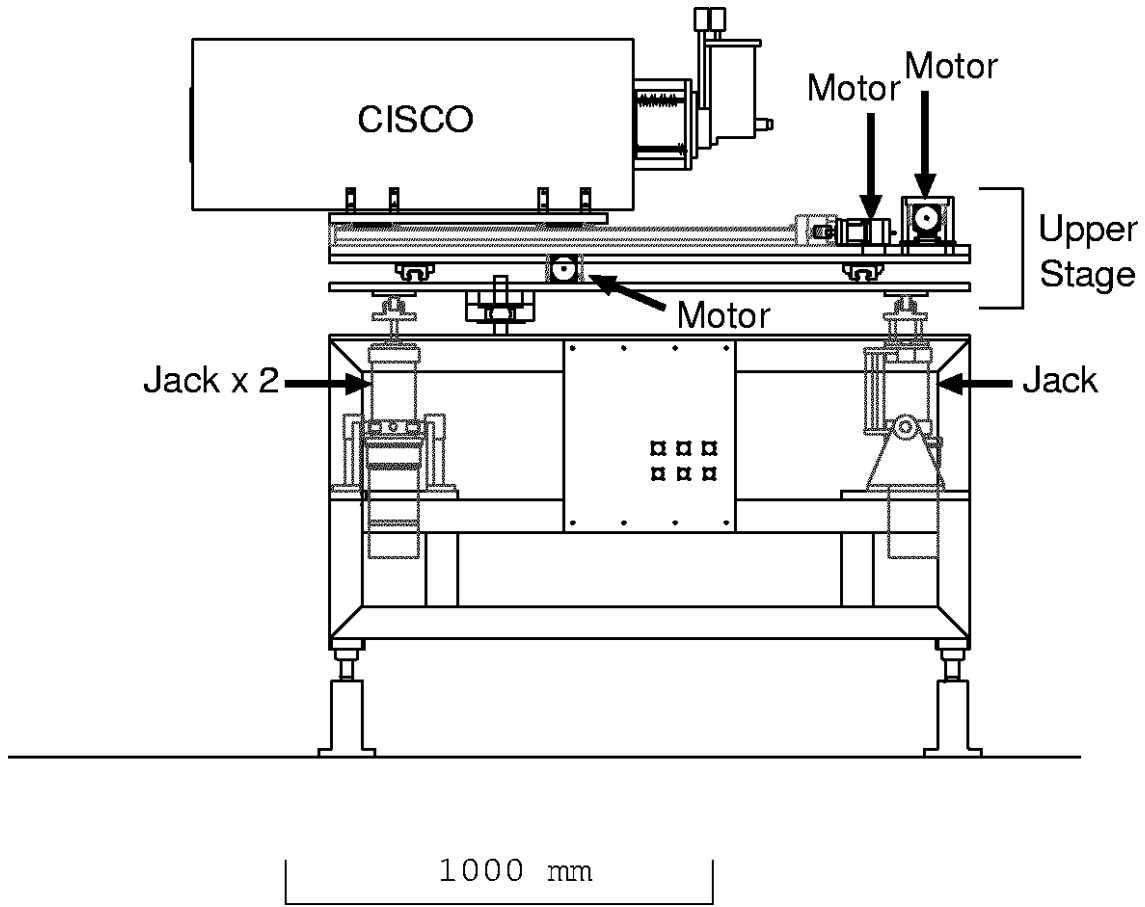}
  \end{center}
  \caption{Schematic view of the CISCO stage.}\label{fig:cstage}
\end{figure}

The cryostat of CISCO is mounted on a support, called the CISCO stage, made
of aluminum-alloy with dimensions of 1500 mm $\times$ 1000 mm $\times$
1300 mm (figure \ref{fig:cstage}).
Its height and inclination are adjusted by three mechanical jacks 
at the bottom. These jacks are controlled through the parallel I/O 
interface of the CIC board in the VME rack.
A movable stage, called the upper stage, is mounted on these jacks.
There are three stepper motors controlled by the 
motor driver boards in the VME rack to adjust the X--Y movement and 
level rotation on the stage.

\subsection{Array Readout Electronics and Data Acquisition}
We exploit a 1024$\times$1024 format HgCdTe array detector, HAWAII, 
manufactured by Rockwell Scientific Co.
The overall layout of the array readout system is illustrated in 
figure \ref{fig:layout}.
It consists of four boards: a fanout board, a preamp board, 
an ADC board, and a clock driver board.

\begin{figure}
  \begin{center}
    \FigureFile(100mm,80mm){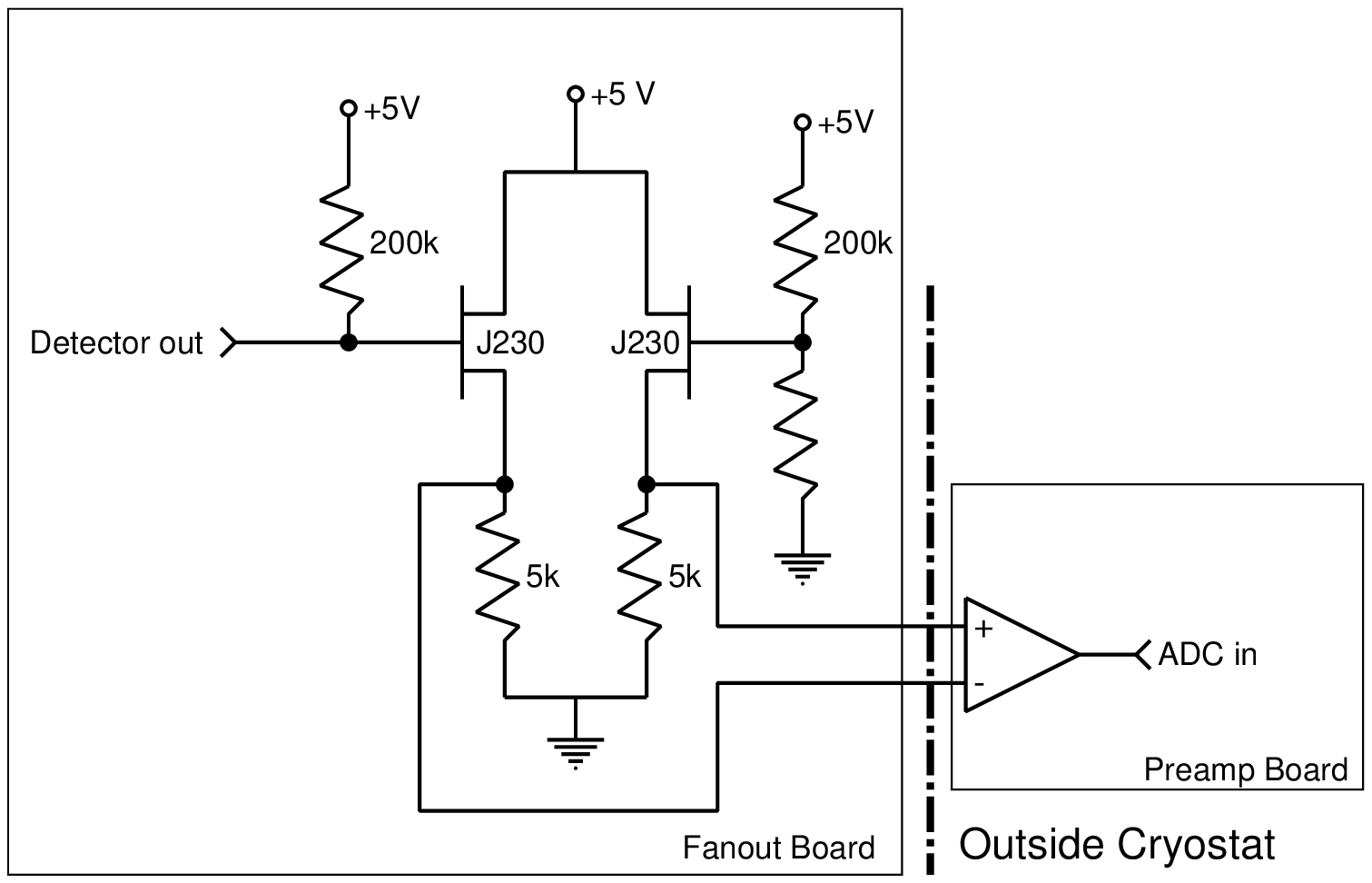}
  \end{center}
  \caption{Schematic of a paired FET buffer on the fanout board.}\label{fig:fanout}
\end{figure}

The HAWAII array has four quadrants, and each quadrant has one output.
There are two channels for each output;
one is buffered by a built-in FET on the multiplexer and the other
is unbuffered. Because strong glow of the built-in FETs are 
is noticed when they are activated (Hodapp et al. 1996), we decided to use
unbuffered outputs and to buffer them by external JFETs on the
fanout board. Figure \ref{fig:fanout} shows a schematic of a JFET buffer 
on the fanout board. It has two n-channel JFETs; one is to buffer the
signal from the array with a gain of 0.82; the other generates
a reference voltage. 
The output is therefore a differential signal which cancels out
noise coming in from the power lines, ground, signal cables and so on.
The outputs are next fed to amplifiers with a gain of 10.0 on the
preamp board, filtered by low-pass filters with a gain of 1.27 and a cut-off
frequency of 1.25 MHz, and digitized by 16 bit A/D converters on the 
ADC board.

Clocks to drive the array and A/D converters are generated on the
clock driver board, using clock patterns created by the CIC board in the VME rack.
The pixel rate for scanning the array is set to 5.2 $\mu$s,
corresponding to a frame rate of 1.4 s.

Acquisition of the digitized data and the generation of clock patterns
to drive the array and A/D converters are carried out by a VME-bus subsystem,
called Messia III (Sekiguchi et al. 1992), 
developed by the National Astronomical Observatory of Japan.
It consists of a VMI board and a CIC board on the VME-bus, an SIF board on the
S-bus of the host workstation, and a software package based on Tcl/Tk
to control these boards.
The VMI board has 32 Mbytes of frame memory to acquire digitized pixel data.
It also works as a VME-bus controller and communicates with the host
workstation through optical fiber via SIF board.
The CIC board has two digital signal processors (DSPs) and generates
clock patterns. It also has various types of I/O interfaces (parallel 
I/O, RS-232C, A/D converter, and D/A converter).

The standard procedure to read out the array follows correlated double
sampling: reset/scan the array, wait for the exposure time, scan the
array again, and subtract two scans to produce a frame.
Due to the design of the multiplexer of the HAWAII array, the first
reset/scan is carried out as follows:
reset one line of pixels simultaneously, scan the line and move to the next
line.
However, it was found out that resetting the line in this
procedure creates offset patterns on a final image.
We therefore decided to insert a reset/dummy-scan before the real scan
to avoid these patterns when the integration time is longer than 10 s,
and use the following procedure;
reset/dummy-scan -- scan -- wait for the exposure time -- dummy-scan -- scan.

It takes 1.4 s to do a single scan, 2 s to transfer the 
image data from the frame memory on the VMI board to the main memory of the
host workstation, and 2 s to convert it into a FITS file.
Conversion to a FITS file being performed as a background
process without taking extra time, overheads to acquire a single frame 
are 4.8 s for an exposure time of over 10 s and 3.4 s under
10.

\subsection{Software}
\begin{figure}
  \begin{center}
   \FigureFile(130mm,80mm){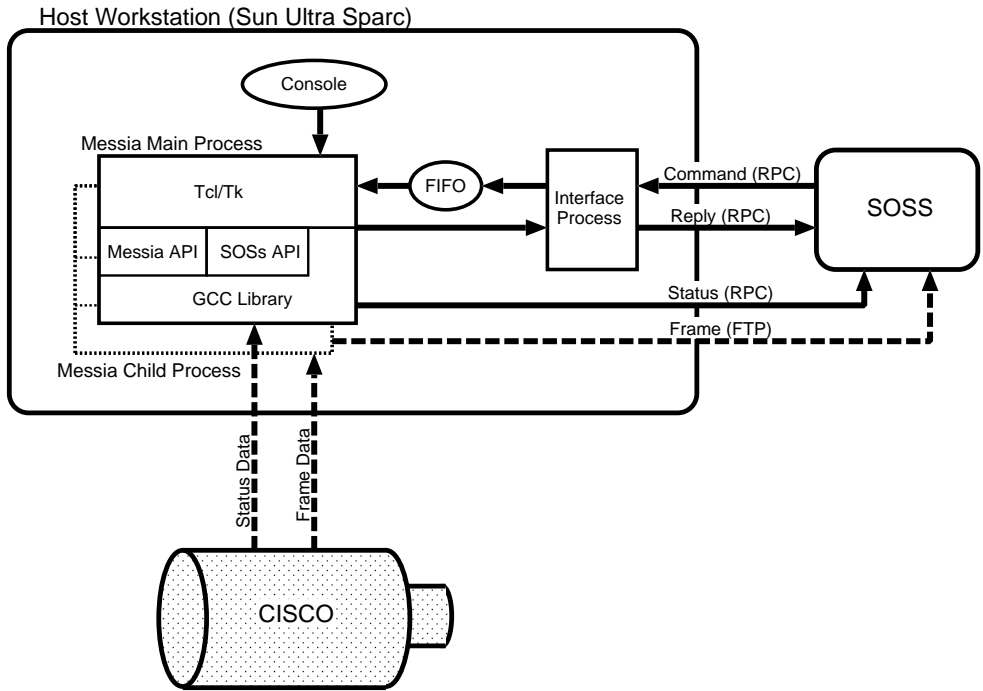}
  \end{center}
  \caption{Block diagram of the software system of CISCO.}\label{fig:software}
\end{figure}

A block diagram of the software system of CISCO is shown in figure
\ref{fig:software}. 

The center of the system is software called ``messia'', 
written in C-language and having a Tcl/Tk interface.
It is based on the libraries of the Messia III package and also employs an
interface library delivered by the software group of the Subaru telescope. 
It can be operated either from the console of the host workstation 
or from the Subaru observation software system (SOSS:
Kosugi et al. 1997).

SOSS is a software system which controls both the telescope and
instruments in close cooperation, and the control-center of the Subaru
telescope. 
A command executed by SOSS is transferred to the host workstation using
a remote protocol (RPC).
It is first received by an interface process, interpreted 
and handed over to messia using FIFO.
Messia forks to create a child process to run the command.
When the command is completed, the main process returns a
signal to the interface process and a completion signal 
is sent back to SOSS.
The frame data acquired is directly transferred to SOSS using FTP.

\section{Performance of the HAWAII Array}
We carried out several tests to evaluate the performance of 
the  HAWAII array and its readout system.
The tests were all carried out with a reset-voltage of 0.5V and 
under a detector temperature of 77 K.

In the following subsections, we call the [ X = 1 -- 512, Y = 1 -- 512 ] region 
of an image a quadrant 1, [ X = 1 -- 512, Y = 513 -- 1024 ] a quadrant 2, 
[ X = 513--1024, Y = 1--512 ] a quadrant 3 and
[ X = 513--1024, Y = 513--1024 ] a quadrant 4. 

\subsection{Conversion Factor}
The conversion factor $g$ $(e^-/\rm ADU)$ is measured from the relation of 
the signal count and the noise, which are given by the photon statistics,
\begin{equation}
 N=\sqrt{N_{\rm read}^2+\frac{S}{g}} ,
  \label{equation:conv}
\end{equation}
where $N$ is the noise, $S$ input signal and $N_{\rm read}$ 
readout noise per frame, all in units of ADU.

To measure the conversion factor, we took images of the thermal emission
from the wall of the Nasmyth stage of the Subaru telescope.
Here, we used $K^{\prime}$ and $\rm H2(2-1)$ filters.
The amount of input signal was adjusted by rotating the filter wheel 
slightly and vignetting the collimated ray.
The exposure time was varied from 5 to 20 s, and two frames were taken 
with the same exposure time.

The noise was calculated by subtracting the 200$\times$200 pixels
region of the two frames, rejecting bad pixels, taking a 
standard deviation of pixel values and dividing it by $\sqrt{2}$. 
However, this method requires flatness of the quantum efficiency, the
dark current and the readout noise.
To evaluate the effect of the flatness of the quantum efficiency, 
we divided the subtracted frame by the flat frame.
We found a negligible difference in the standard deviation value, meaning that 
the difference in the quantum efficiency between the pixels
does not contribute to the noise.
The effect of the flatness of the dark current and the readout noise was
also found to be negligible by comparing the thus-calculated noise of
dark frames with that obtained by an ordinary procedure, as 
used in the next subsection.

\begin{figure}
  \begin{center}
    \FigureFile(90mm,80mm){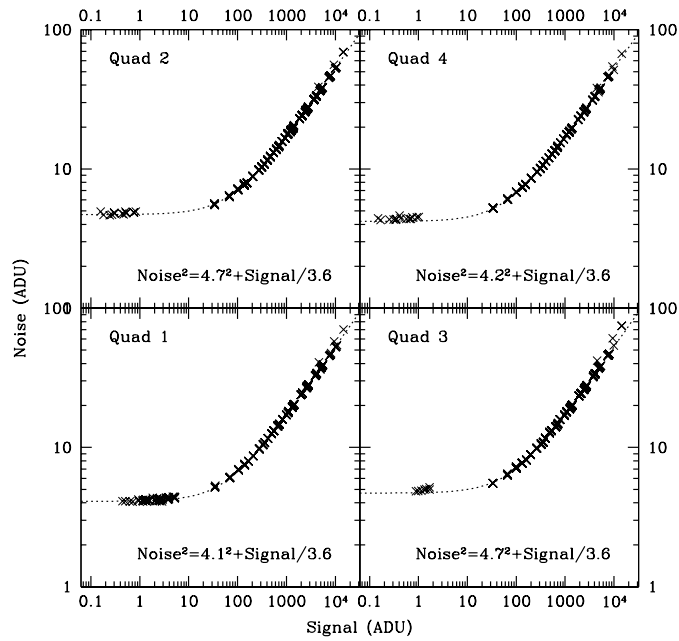}
  \end{center}
  \caption{Relationship between the signal and the noise in the ADU unit.}\label{fig:convfactor}
\end{figure}

The relationship between the signal and the noise of $200\times200$ pixels
region in each quadrant is shown in figure \ref{fig:convfactor}.
We fit these plots by equation (\ref{equation:conv}) and 
determined the conversion factor to be 3.6 $e^-/\rm ADU$ in all four
quadrants.
This value matches well with the estimated value of 2.5 -- 4.5 $e^-/\rm ADU$, 
calculated using the measured gain of readout 
electronics and the assumed parameters of the array detector taken from
the data-sheet provided by the manufacturer.

In the following discussion, we assume the conversion factor to be
3.6 $e^-/\rm ADU$.

\subsection{Readout Noise}
To measure the readout noise, we obtained 10 dark frames of 60 s exposure.
The filter wheels were set to ``blank'', 
which is a combination of $\rm H2(1-0)$ and $\rm N204$,
and achieved sufficient darkness.
The number of multiple samples $( n )$ was varied from 1 to 12.
Here, an $n$-times multiple sample means:
reset/dummy-scan -- scan $n$ times -- wait for the exposure time --
dummy-scan -- scan $n$ times.
We created standard deviation frames from the dark frames and defined
the readout noise to be a mean value of a $128\times128$ pixels region
with a low dark current in each quadrant of the standard deviation frames.

\begin{figure}
  \begin{center}
    \FigureFile(90mm,80mm){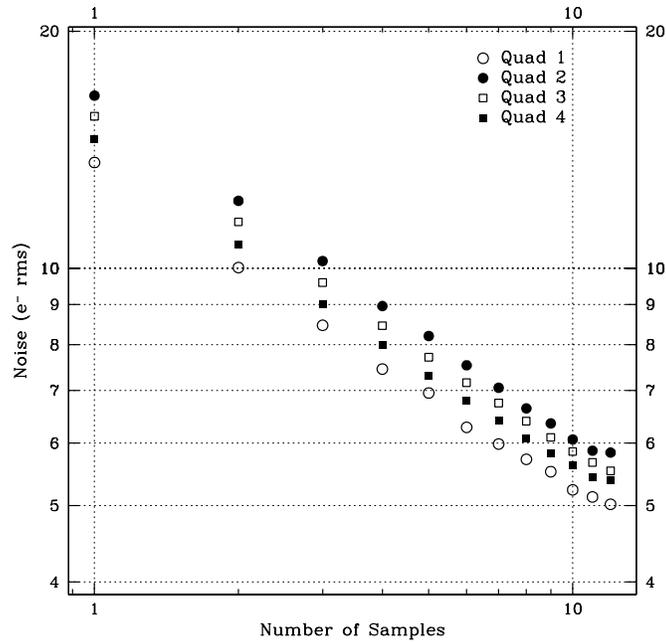}
  \end{center}
  \caption{Relationship between the readout noise and the number of samples.}\label{fig:multinoise}
\end{figure}

We show the relationships between the number of multiple samples $(n)$ and the
readout noise in figure \ref{fig:multinoise}.
The readout noise was found to be 14 -- 17 ($e^-$ r.m.s/frame) 
with a single correlated double sampling, 6.3 -- 7.5 ($e^-$ r.m.s/frame) 
with 6-times multiple sampling and 
 5.0 -- 5.8 ($e^-$ r.m.s/frame) with 12 times multiple sampling.
These values are small enough to reach the background-limited noise condition
even under a low background level using OHS.

\subsection{Linearity and Full Well}
To measure the linearity and the full well, we took images of the thermal
emission from the wall of the telescope.
We used $\rm N204$, $\rm H2(1-0)$ and $\rm H2(2-1)$ filters.

The exposure time was set to 750 s and non-destructive readouts
were made every 30 s during exposures to measure the amount of
collected electrons. 
A total of 26 frames were acquired per exposure.
Two exposures were carried out for each filter to 
check the stability of the input flux.

\begin{figure}
  \begin{center}
    \FigureFile(90mm,80mm){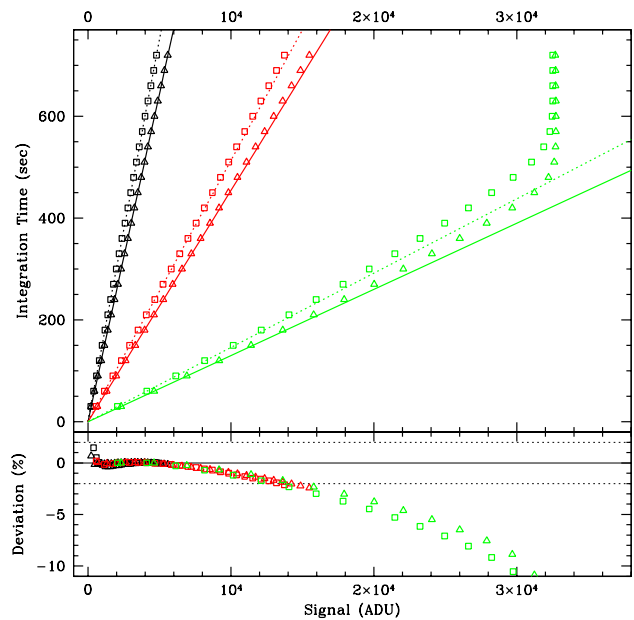}
  \end{center}
  \caption{Linearity of the HAWAII array. The open triangles represent the
 $50\times10$ pixels of higher quantum efficiency, while the open squares
 represent those of lower quantum efficiency. The upper box shows the relationship between the
 exposure time and the signal. The fitted lines in the range from
 100 to 5000 ADU are overlapped. The lower box shows the deviations from
 the fitted lines. The black points were taken with the $\rm N204$ filter,
 the red points with $\rm H2(1-0)$, and the green points with $\rm H2(2-1)$. }\label{fig:linearity} 
\end{figure}

The upper box of figure \ref{fig:linearity} shows the relationship between 
the integration time and the signal.
The open triangles represent a $50\times10$ pixel regions with a higher 
quantum efficiency, while the open squares represent that with a lower quantum
efficiency.

We fitted these plots with lines in the range from 100 to 5000 ADU; 
residuals from the fitted lines are shown in  the lower box
of figure \ref{fig:linearity}.
The deviations from the lines are smaller than 2\% below 14000 ADU.
The signals are saturated at 32500 ADU ($1.17\times 10^{5}$ $e^-$)

\subsection{Dark Current}
To obtain dark-current frames, both slit-units were closed and 
the filter wheels were set to blank.
Ten 3200 s exposures were carried out.
During the exposures, a non-destructive readout was 
performed every 160 s and 21 frames were acquired in total.

To avoid the effect of the reset anomaly of the array (Finger et al. 2000), 
we discard the first 11 frames of each integration set.
Two 1280 s exposure frames were created using the 
12th, 13th, 20th, and 21st frames of each set 
and combined to produce a final dark-current image.

\begin{figure}
  \begin{center}
   \FigureFile(110mm,80mm){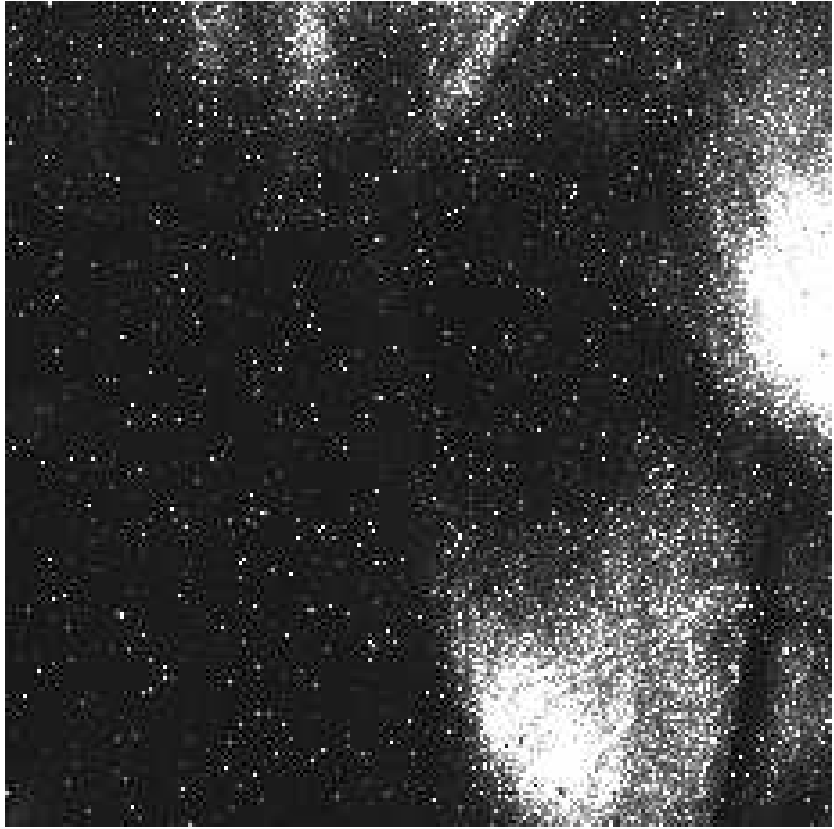}
  \end{center}
  \caption{Image of the dark current. There are two ``hot'' regions which
 show a 100-times higher dark current than the other.}\label{fig:darkimage}
\end{figure}

\begin{figure}
  \begin{center}
    \FigureFile(90mm,80mm){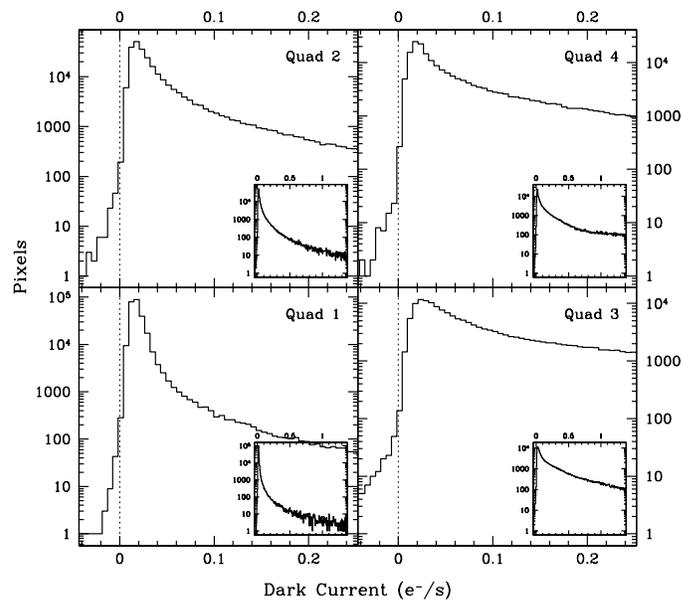}
  \end{center}
  \caption{Histograms of the dark current in each quadrant of the
 HAWAII array.}\label{fig:darkhist}
\end{figure}

We show the dark current image in figure \ref{fig:darkimage}.
The existence of ``hot'' regions where the dark current exceeds
1$e^-/\rm s$ 
can be seen in quadrants 3 and 4.
However, the overall value of the dark current is fairly low.
Figure \ref{fig:darkhist} shows a histogram of the dark-current frame.
The mode value of each quadrant was found to be 0.015 -- 0.03 $(e^-/\rm s)$.

\subsection{Variation of the Quantum Efficiency}
\begin{figure}
  \begin{center}
   \FigureFile(110mm,80mm){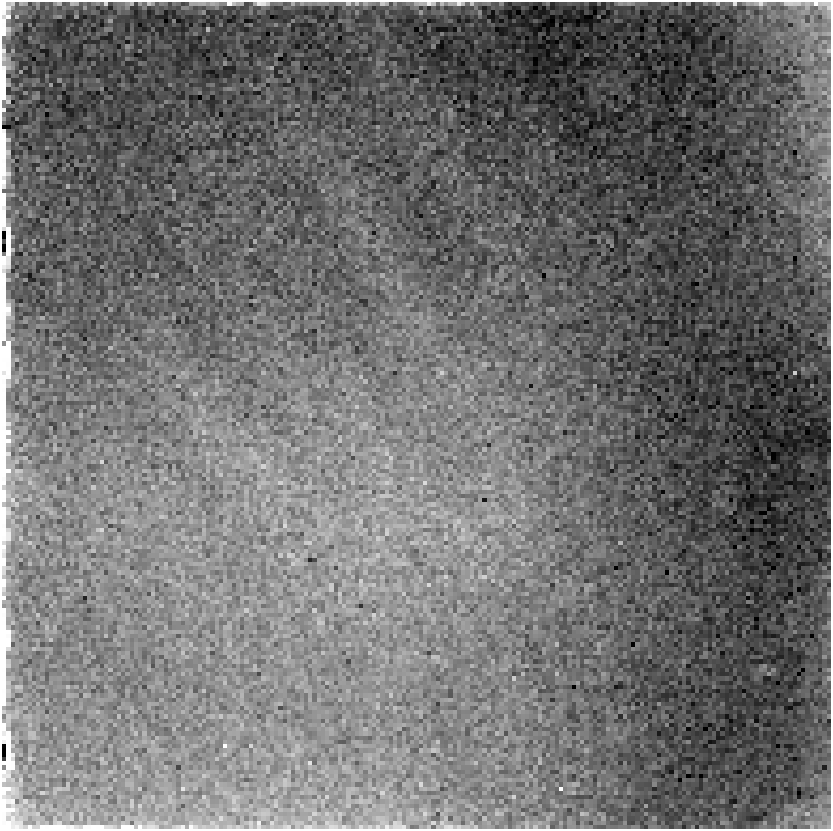}
  \end{center}
  \caption{Image of the sky-flat of the $K^{\prime}$-band, taken at 
 the Nasmyth focus of the Subaru telescope.}\label{fig:flatimage}
\end{figure}

\begin{figure}
  \begin{center}
    \FigureFile(90mm,70mm){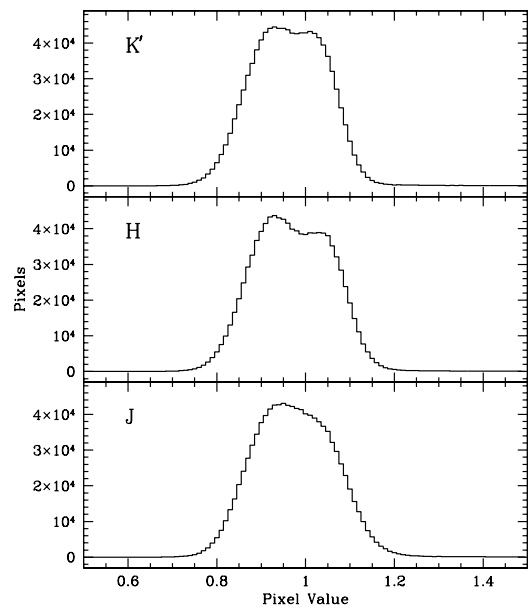}
  \end{center}
  \caption{Histograms of the sky-flat frames of the $J$-, $H$- and
 $K^{\prime}$-band.}\label{fig:flathist}
\end{figure}

Flat-field images were produced using sky images
of deep imaging observations.
Figure \ref{fig:flatimage} shows the $K^{\prime}$-band flat 
image, and figure \ref{fig:flathist} shows
histograms of the flat images of the $J$-, $H$- and $K^{\prime}$-band.
All of the images were normalized by the central $512\times512$ pixels.
The flatness of the quantum efficiency was good, and the
standard deviations of the histograms were measured to be 
0.088, 0.086 and 0.081 for the $J$-, $H$- and $K^{\prime}$- bands,
respectively.

\subsection{Excess Dark Current}

When a pixel of the HAWAII array is illuminated by relatively strong
radiation, it is known that it shows an excess dark current in the subsequent
exposure (Hodapp et al. 1996; Finger et al. 1998).
This phenomenon becomes distinct when the sky-background level
increases or decreases, i.e. when the filter is changed, or the
observation mode is changed from imaging to spectroscopy.

We evaluated the effect of the excess dark current as follows.
First, we set the filter to $K^{\prime}$ and took a single ``bright'' frame of 
thermal emission from the wall of the telescope.
One unit of the slit blades was closed halfway to make a dark region, 
which was used to compensate for the offsets caused by a reset anomaly.
The exposure times for the bright frames were 20 s and 5 s, and the 
input signal were 16000 ADU and 5000 ADU, respectively.
After the exposure, we set the filter to blank and took series of dark frames.
To evaluate the effect of resetting the pixels, two exposure times for
the dark frames were used, the longer ones being 100 s and the shorter
ones being either 10 sec or 12.5 s. 

\begin{figure}
  \begin{center}
    \FigureFile(130mm,90mm){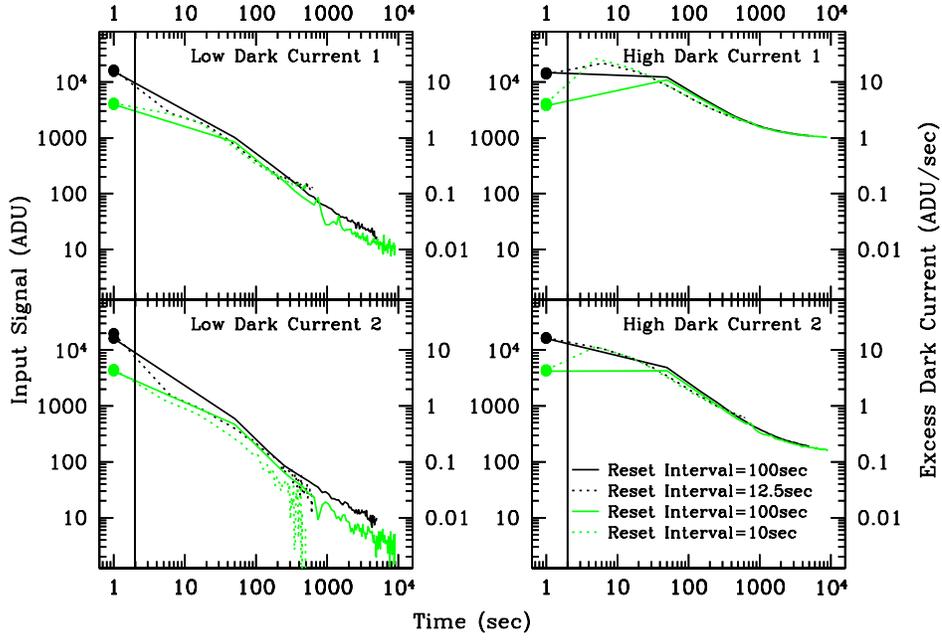}
  \end{center}
  \caption{Excess dark current of four different $100\times100$ pixels
 regions. The filled circles on the left of each plot represent the input
 signals of the bright frames.}\label{fig:latent}
\end{figure}

We measured the excess dark current in four $100\times100$ pixel
regions; two with a low dark current ($<0.1\,e^-/\rm s$) and two with a high
dark current ($>0.1\,e^-/\rm s$).
Figure \ref{fig:latent} shows the relation between the excess dark current and
time. The filled circles at the left edge of each plot represent the input
signal of the bright frames. 
It can be seen that even the input flux varies by more than three times, 
the amount of the excess dark current changes by only less than 20\%.
Also, decay time-scale of the excess is slightly affected 
by the change of either the input flux or the reset interval.

In a normal procedure of spectroscopic observation, we first do imaging 
to introduce the target onto the slit where the background level is a few
$\times$ 1000 to 10000 ADU.
The above-mentioned result suggests that after slit-introduction of the target,
we should wait for more than 200 s for the excess dark current 
to decay below a level that would not affect the spectroscopic observation
(0.1 $e^-/\rm s$).

\section{System Performance}
CISCO was first tested on the 1.5 m telescope at the campus of the 
National Astronomical Observatory of Japan from 1997 November to 
1998 May.
It was then transported to the Subaru telescope,
and mounted on the Cassegrain focus
to obtain astronomical first-light in 1999 early January.
After one year of test observations, CISCO was relocated to the
Nasmyth focus in 1999 December, and has been used as either a back-end
camera for OHS or a general-use infrared camera/spectrograph. 
In this section, we present results concerning the system performance 
using the data acquired at the Nasmyth focus.

\subsection{System Efficiency}
The system efficiency is measured by the obtained frames of
the UKIRT faint standard stars (Hawarden et al. 2001).
The measured efficiencies are
22\% in $J$, 32\% in $H$ and 36\% in $K^{\prime}$ and $K$.
They include the atmospheric transmittance, the optical throughput and
the quantum efficiency of the array detector. 

\subsection{Limiting Magnitudes}
From deep imaging observations, the background brightness was estimated to be
15.2 mag in $J$, 13.6 mag in $H$, and 13.1 mag in $K^{\prime}$, although 
they fluctuate by more than 0.3 mag according to night and time.
From these values, we then estimated the limiting magnitudes for the
imaging, which are 23.5 mag in $J$, 22.9 mag in $H$, and 22.4 mag in
$K^{\prime}$. The limiting magnitudes for the spectroscopy
were also estimated. All of these results are summarized in table 3.

Several scientific programs were conducted using 
CISCO during the commissioning phase of the Subaru telescope for roughly
one year.
One of such programs was a very deep survey in the $J$- and $K^{\prime}$-
bands, devoting more than 20 nights. It is noteworthy that the resultant
performance in terms of the limiting magnitude is slightly higher than our
estimates, because observations for the deep imaging were made only under
the best seeing conditions (Maihara et al. 2001). The overall seeings
reported in the paper were \timeform{0.35"} in FWHM in the
$K^{\prime}$-band, and \timeform{0.45"} in the $J$-band.
In any case, these match well with the results of real imaging
observations.

\section{Summary}
We have developed a near-infrared camera and spectrograph, CISCO.
CISCO was originally designed as a back-end camera for an OH-airglow
suppressor spectrograph for the Subaru telescope, and also to be used
as an ordinary infrared camera directly attached to the Nasmyth focus of the
telescope. It is capable of \timeform{1.8'}$\times$\timeform{1.8'} field-of-view
imaging or long-slit spectroscopy from 0.9 to 2.4 $\mu$m.

The performance of the HAWAII array was evaluated. The readout noise and the
dark current were shown to be small enough to be operated under a low-background
condition using OHS. The effect of the excess dark current is found to
last for a long time, requiring more than a few hundred seconds to decay.
The total system performance was also evaluated using the 
observational data at the Subaru telescope.
\\[6pt]

We are indebted to all staff members of the Subaru telescope, NAOJ for
support during our testing and observational runs.
We thank K. Hodapp and the electronics staff in IfA, University of Hawaii
for providing us with a specific design of the readout electronics for HAWAII,
M. Tanaka, Y. Kobayashi, T. Pyo, and T. Sekiguchi for
supporting out test observations at National Astronomical Observatory of
Japan.
We also thank A. T. Tokunaga for useful comments and suggestions.
K.M. was supported by a Research Fellowship for Young
Scientists of the Japan Society for the Promotion of Science.
This work was supported by a Grant-in-Aid for Scientific Research (B),
Japan (No. 11440065).


\onecolumn

\begin{table}
  \caption{Parameters of optical elements of CISCO.}\label{tab:opt}
  \begin{center}
    \begin{tabular}{p{10pc}cc}
     \hline
     \hline
     Element & Size(mm) & Quality\\
     \hline
     Entrance window\dotfill & 150 &$\rm CaF_2$\\
     Field lens\dotfill & 120 & $\rm CaF_2$\\
     Collimator lens 1\dotfill & 90 & $\rm SiO_2$\\
     Collimator lens 2\dotfill & 80 & $\rm CaF_2$\\
     Collimator lens 3\dotfill & 80 & $\rm SiO_2$\\
     Ly\^ot stop\dotfill & 24 & \\
     Refocus lens 1\dotfill & 50 & $\rm BaF_2$\\
     Refocus lens 2\dotfill & 50 & MgO\\
     Refocus lens 3\dotfill & 50 & $\rm Ca_2$\\
     Refocus lens 4\dotfill & 50 & $\rm BaF_2$\\
     Refocus lens 5\dotfill & 50 & $\rm SiO_2$\\
     \hline
    \end{tabular}
  \end{center}
\end{table}

\begin{table}
  \caption{Filters, grisms and a prism contained in the filter wheels.}\label{tab:filter}
  \begin{center}
    \begin{tabular}{p{6pc}ccl}
     \hline
     \hline
     Name & Wavelength coverage(\micron) &Size(mm) & Remarks\\
     \hline
     $K$\dotfill & $2.03-2.37$ & 37$\phi$& Broad-band filter\\
     $K^{\prime}$\dotfill & $1.95-2.29$ & 37$\phi$& Broad-band filter\\
     $H$\dotfill & $1.49-1.78$ & 37$\phi$& Broad-band filter\\
     $J$\dotfill & $1.17-1.33$ & 37$\phi$& Broad-band filter\\
     $z$\dotfill & $0.90-1.10$ & 37$\phi$& Broad-band filter\\
     \hline
     N204\dotfill & $2.023-2.043$ & 37$\phi$& Narrow-band continuum\\
     H2(1-0)\dotfill & $2.110-2.130$ & 37$\phi$& $\rm H_2(1-0)$ emission-line\\
     N215\dotfill & $2.137-2.158$ & 37$\phi$& Narrow-band continuum\\
     H2(2-1)\dotfill & $2.239-2.261$ & 37$\phi$& $\rm H_2(2-1)$
     emission-line\\
     \hline
     $zJ$Gr \dotfill & $0.88-1.36$ & $31\times31$& Grism, 200 lines/mm\\
     $JH$Gr \dotfill & $1.11-1.81$ &  $31\times31$& Grism, 285 lines/mm\\
     $K$Gr \dotfill & $1.88-2.42$ &  $31\times31$& Grism, 165 lines/mm\\
     Pr \dotfill & $0.80-2.50$ &  $31\times31$& Zenger prism\\
     \hline
    \end{tabular}
  \end{center}
\end{table}

\begin{table}
  \caption{System performance of CISCO.}\label{tab:performance}
  \begin{center}
    \begin{tabular}{p{18pc}p{6pc}p{6pc}p{6pc}}
     \hline
     \hline
     & $J$ & $H$ & $K^{\prime}$ \\
     \hline
     Wavelength ($\mu$m)\dotfill & 1.25 & 1.64 & 2.13\\
     Bandwidth ($\mu$m)\dotfill & 0.16 & 0.28 &0.34\\
     Efficiency (mag/$\rm e^-$)\dotfill & 26.4 & 26.5 & 26.1\\
     Background brightness (mag arcsec$^{-2}$)\dotfill & 15.2 & 13.6 & 13.1\\
     Limiting magnitude$^*$ (mag)\dotfill & 23.5 & 22.9 & 22.4\\
     Limiting magnitude$^{\dagger}$ (mag)\dotfill & 21.2 & 20.2 & 19.7\\
     \hline
     \end{tabular}
  \end{center}
\vspace{6pt}
\par\noindent
 *~ Imaging, S/N=5, 3600 s integration, \timeform{1''} aperture.\\
 $\dagger$~ Spectroscopy, S/N=5, 3600 s integration, \timeform{1''}
 slit.
\end{table}

\clearpage

\end{document}